\documentstyle[preprint,aps]{revtex}%
\begin{document}
%
\preprint{WM--95--108, UVA-INPP-95/10, CEBAF-TH-95--06}
%
%
\begin{title}
{\bf Extraction of the Ratio of the
Neutron to Proton Structure Functions from Deep Inelastic Scattering}
\end{title}
\author{S.Liuti $^*$}
\address
{Continuous Electron Beam Accelerator Facility  \\
12000 Jefferson Ave, Newport News, Virginia 23606, USA}
\author{Franz Gross}
\address
{College of William and Mary, Williamsburg, VA 23185, USA\\
and\\
Continuous Electron Beam Accelerator Facility  \\
12000 Jefferson Ave, Newport News, Virginia 23606, USA}
\date{\today}
\maketitle
\begin{abstract}
We study the nuclear ($A$) dependence of the European Muon Collaboration
(EMC) effect at high values of $x$ ($x \geq 0.6$).
Our approach makes use of conventional
nuclear degrees of freedom within the Relativistic
Impulse Approximation.
By performing a non-relativistic series expansion
we demonstrate that relativistic corrections
make a substantial contribution to the effect at $x \gtrsim 0.6$ and
show that the ratio of neutron to proton
structure functions extracted from a global fit to
all nuclei is not inconsistent with values obtained from the deuteron.
\end{abstract}


\newpage

The famous high energy deep inelastic lepton scattering results
obtained in 1983 at CERN
by the European Muon Collaboration (EMC) \cite{EMC} and at
SLAC in 1984 \cite{Arnold} (now known as the EMC effect) showed that
quark momentum distributions are modified in the
nuclear medium.  The scattering
is well described by assuming that the scattered lepton
interacts with a bound quark by exchanging a virtual quanta
with four-momentum $Q^2=q^2-\nu^2$, where $q=|{\bf q}|$ and $\nu$ are the
three-momentum and energy carried by the quanta, respectively.
The inclusive cross section depends on the Bjorken variable
$x=Q^2/(2M_N\nu)$, which is identified with the fraction of the
total longitudinal momentum carried by the struck quark.
In this Letter we will discuss recent progress in the interpretation of
the data for large quark momentum fractions $x\gtrsim 0.6$.

Figure 1 shows the ratio $R_A(x)$ of the nuclear structure function per
nucleon,
$F^A_2(x)/A$, to the deuteron structure function per nucleon,
$F^D_2(x)/2$.
The data are for an iron nucleus, but the results are similar for all nuclei
with mass number $A\geq 3$.  If the nuclear medium had no effect on the
quark momentum distribution, the ratio would be unity; the
(up to 20\%) deviations
of the nuclear structure functions are direct evidence for the effect of the
nuclear medium.

Throughout the years, a variety of models have been proposed to explain the
EMC effect at large $x$ (for a review see  \cite{Berger}).  Some invoke
additional (sometimes exotic) components of the nuclear wave function, while
other, more conventional models describe the EMC effect in terms
of nucleon binding.  In this latter picture,
the rise above one as
$x \to 1$
is due to Fermi motion.  In Ref.~\cite{Liu1}, referred to as CL in
this paper, it was pointed out that a
nonrelativistic calculation based on binding plus short range
nucleon-nucleon ($NN$) correlations (generated by the $NN$ potential)
were sufficient to account for most of the
effect at $x<0.5$, but the result for large $x$ (the long dashed line
in Fig.~1) was much too small.  As this  model requires only a very
small number of
parameters which can be determined to high precision from other data
\cite{BenPan}, agreement
cannot be obtained by adjusting the parameters, leaving open the possibility
that exotic components could play a role in the explanation of the
effect \cite{Liu2}.

In Ref.~\cite{GroLiu}, which we will refer to as GL, we introduced a new
relativistic formulation of the impulse approximation (which we refer to
as the RIA), based on the relativistic spectator model \cite{Gross}.
Using this formulation, we found that we can explain the
EMC effect in the region of large $x$ entirely in terms of conventional
nuclear degrees of freedom, as shown in Fig.~1.

In this Letter we reinforce these conclusions by showing that (i) the
theoretical uncertainty of the RIA calculation is not large, so that the
agreement presented in GL is no accident, (ii) the
relativistic effects can be rather simply parameterized, and hence their
physical origin easily understood, and (iii) the new
relativistic theory removes a descrepancy between the
results obtained for the neutron structure function, $F_{2n}(x)$, in two
different ways: from measurements of the deuteron, and from global fits
to nuclei with all values of $A$.

To begin the discussion, we recall that
the $A$ dependence of the experimental ratio $R_A(x)$
has been generally parametrized as a product
of a function of $x$ times a function of $A$:
\begin{equation}
R_A(x) -1 \simeq \alpha(x) \beta(A) \, . \label{eq1}
\end{equation}
This is referred to as ``factorization''.  Note that the parameterization
\cite{Arnold}
$R_A(x) \propto A^{\alpha(x)}$ takes this form if the exponent $\alpha(x)$
is sufficiently small.
A factorized form for $R_A(x)$ can also be obtained
from a naive nonrelativistic IA description.
In this case the structure function, $F_2^A$, is given
by a linear convolution formula \cite{Aku,FSPRep}
\begin{equation}
\left[ F_2^A(x) \right]_{IA} = \int_x^\infty dz f_A(z) F_2^N(x/z)\, ,
\label{IA}
\end{equation}
where $F_2^N$ is the structure function for an
off-shell nucleon (which is assumed to have the same
form as the on-shell one), $z =A(k\cdot q)/(P_A \cdot q) \simeq k^+/M_N$
is the light-cone momentum
fraction of the struck
nucleon with four-momentum $k= (k_o,{\bf k})$,
$M_N$ and $M_A$ are
the masses of the nucleon and nucleus $A$, respectively, and $f_A(z)$ is
the nucleon light cone momentum distribution.

To obtain the factorized form (\ref{eq1}) from the convolution formula
(\ref{IA}), it is useful to exploit the fact
that the nuclear momentum distribution $f_A(z)$ is sharply peaked
around $z=1$, and expand the factor $F_2^N(x/z)$ in Eq.~(\ref{IA}) in
powers of $(1-z)$ around $z=1$
\begin{equation}
F_2^N(x/z) = F_2^N(x) + x\frac{\partial F_2^N(x)}{\partial x}\,(1-z)
+{1\over2}\left[x^2 \frac{\partial^2 F_2^N(x)}{\partial x^2}+
2x\frac{\partial F_2^N(x)}{\partial x} \right] (1-z)^2 +\cdots\,  .
\end{equation}
The coefficients of the expansion, accurate to order $1/M_N$, are
therefore
\begin{eqnarray}
c_0&&= \int_x^\infty dz f_A(z) \simeq \int_0^\infty dz f_A(z)\simeq A
\nonumber\\
c_1&&= \int_x^\infty dz f_A(z) (1-z) \simeq A\left[ {\langle E \rangle
\over M_N}
-{2\over3}{\langle T\rangle\over M_N} \right]\nonumber\\
c_2&&= \int_x^\infty dz f_A(z) (1-z)^2\simeq
{2\over3}\,A\,{\langle T\rangle\over M_N} \, , \label{coeffs}
\end{eqnarray}
where $\langle T \rangle$ is
the average kinetic energy of the nucleon in the nucleus and $\langle E
\rangle =\langle M_{A-1}\rangle+M_N - M_A$ is the
average removal energy, with $\langle M_{A-1}\rangle$ the average mass of
the spectator
$A-1$  nuclear system (in this discussion we neglect the recoil energy of
the $A-1$ system).  Details of the derivation of these
coefficients (\ref{coeffs})
are discussed in CL.  The coefficient $c_0$ is
just the normalization of the light cone momentum distribution, and
$c_1$ and $c_2$ can be related to $\langle T \rangle$ and $\langle E\rangle$
by exploiting the connection between $f_A(z)$ and the nucleon
three-momentum distribution, $n_A(k)$
\begin{eqnarray}
f_A(z) &&= 2\pi \, z \int_0^\infty dk \, k \, n_A(k) \int_{-k}^k dk_{_{||}}\,
\delta\left(1-z-{\langle E\rangle\over M_N} -{k_{_{||}}\over M_N}\right)
\nonumber\\
&&=2\pi M_N\; z \int_{k_{min}(z,\langle E \rangle)}^\infty
dk \, k \, n_A(k) \, ,
\label{fzia}
\end{eqnarray}
where $k=|{\bf k}|$ is the magnitude of the three momentum of
the struck nucleon, and $k_{_{||}}$ its component
in the direction of the ${\bf q}$.

Note the presence of the the factor $z$ (sometimes referred to
as the {\em flux factor})
in these equations. This quantity was omitted from some early papers
on nuclear deep inelastic scattering because incorrect assumptions
were made in connecting the relativistic formalism with the
nonrelativistic distributions actually used in the calculations.
Its effect on nuclear structure functions was emphasized in
\cite{FSPRep}, but here we wish to emphasize that the flux factor does not
change the normalization by more than a few percent, which is
not numerically significant in the discussion of the EMC effect
at large $x$, and it has no effect on the coefficient $c_2$.  The
principal effect of the flux factor is to add the
term $-2\langle T\rangle/(3 M_N)$ to the $c_1$  coefficient
in Eq.~(\ref{coeffs}).  This
has a pronounced effect, reducing the size of this coefficient by almost
a factor of two, and
decreasing the size of the EMC effect predicted by the nonrelativistic
impulse approximation (for a detailed discussion, see CL).

A factorized equation of the form (\ref{eq1}) can be obtained by
using the energy-weighted sum rule \cite{Koltun}
\begin{equation}
\langle E\rangle ={A-2\over A-1}\langle T\rangle +{2\epsilon}
\approx {3\over2} \langle T\rangle
\, , \label{Ksum}
\end{equation}
where $\epsilon$ is the binding energy per nucleon for nucleus $A$, and
the second expression uses the approximation $2\epsilon\approx
\langle T\rangle /2$.  With these approximations we obtain
\begin{equation}
\alpha(x)= {3\over2} x \frac{\partial F_2^N(x)}{\partial x}
+ \frac{1}{3} x^2 \frac{\partial^2 F_2^N}{\partial x^2}\, ,\qquad\quad
\beta(A)=\langle T\rangle/M_N \, .  \label{ab}
\end{equation}
Therefore, in nonrelativistic IA,
the $A$ and $x$ dependencies of $R_A-1$ factor, with the $A$ dependence
given entirely by the average kinetic energy of a nucleon in a nucleus,
and the $x$ dependence contained entirely in the term which depends on
derivatives of the free structure function $F_2^N$.
In CL it was shown that
realistic momentum distributions including
$NN$ correlations yield
values of $\langle T \rangle$ which are large enough
to reproduce the EMC effect at $x<0.5$, but the expansion (\ref{ab})
does not explain the behavior of $R_A$ at higher $x$ (as already shown
in Fig.~1).

Calculations performed with the Relativistic Impulse Approximation (RIA)
introduce some important corrections to the IA convoution formula,
Eq.~(\ref{IA}).
In RIA, the nuclear structure function becomes
\begin{equation}
\left[ F^A_2(x) \right]_{RIA} =
\int_x^\infty dz f^{RIA}_A(z,x) ,
\label{RIA}
\end{equation}
with
\begin{equation}
f^{RIA}_A(z,x) =
2 \pi M_N\, z
\int ^\infty _{k_{min}(\langle E \rangle _A,z)} dk\,k\,
n_A(k) \widetilde{F}^N_2(y,k)\,  , \label{ria2}
\end{equation}
where $y=\eta x/z$ ($\eta = A M_N/M_A\simeq 1$)
is the momentum fraction carried by a quark inside the bound nucleon,
$\widetilde{F}^N_2(y,k)$ is the structure function for the bound
(off-shell) nucleon which depends on the longitudinal and transverse
momentum of the nucleon through $z$ and $k$, and $n_A(k)$ can be related
to a covariant nuclear spectral function (which is not known, but is
determined by its relation to $n_A(k)$).  Further analysis of the
relativistic kinematics permits us to express the off-shell
nucleon structure function, $\widetilde{F}^N_2(y, k)$, as a product
of a relativistic phase space factor, $P(y,y')$ times
the on-shell structure function of a shifted argument, $F^N_2(y')$,
\begin{equation}
\widetilde{F}^N_2(y,k) = P(y,y') F^N_2(y')
\label{transformation}
\end{equation}
where $P(y,y')$ describs the phase space of the
spectator quarks (and satisfies the condition $P(y,y)=1$),
and $y'$ is the value of $y$ shifted by the relativistic kinematics:
\begin{mathletters}
\begin{eqnarray}
&&y'  =  y\left(1 -  {\scriptstyle{1\over2}}\Delta\right) +
\left\{\sqrt{\left(b^2(y)+ {\scriptstyle{1\over2}}y\Delta\right)^2 +
y(1-y)\Delta} -b(y)\right\}\, , \label{y'} \\
&&b(y) = {1\over2}
\frac{m_X^2/M_N^2}{1-y} - {1\over2} (1-y) \label{a}\\
&&\Delta = \frac{m^2- k_\mu^2}{M_N^2} =  1-{\left(M_A-M_{A-1}\right)^2\over
 M_N^2} +2{M_A\left(E_{A-1}-M_{A-1}\right)\over M_N^2 }\, ,
\label{delta}
\end{eqnarray}
\end{mathletters}

\noindent where $m_X>M_N$ is the mass of the spectator quarks with
relativistic phase space $P(y,y')$, $m$ is the mass of the struck quark,
$k_\mu^2= k_0^2-k^2\ne m^2$ is the
square of the four-momentum of the off-mass-shell struck quark, and
$M_{A-1}$
is the mass of the recoiling $A-1$-nuclear system.  Note that $\Delta$
is a measure of how far the struck quark is off-shell, and is zero if
it is on-shell.

This important connection between the off-shell and on-shell
nucleon structure function was first obtained in GL.
The most significant difference between the non-relativistic convolution
formula, Eq.~(\ref{IA}), and its
relativistic counterpart, Eq.~(\ref{RIA}),
is the explicit dependence
of the nucleon structure function, $\widetilde{F}^N_2$,
on $\Delta$; when $\Delta=0$ one can easily see that
$y'=y$ and the IA convolution formula, Eq.~(\ref{IA}), is recovered.
We emphasize that Eqs.~(\ref{ria2})--(\ref{delta}) have been derived
from considerations of relativistic kinematics only; any dynamical
dependence of $\widetilde{F}^N_2$
on the nuclear medium, i.e. an explicit dependence on $k_\mu^2$ not due
to relativistic kinematics, has been disregarded.  This means that our
relativistic formulae are truly consequences of the assumption that the
nucleons are not modified by the nuclear medium.

Both the $A$ and $x$ dependence of the EMC-effect
at large $x$ is significantly altered by the parameter $\Delta$,
which includes the relativistic corrections to the IA.
We can easily display (approximately) the
effect of these relativistic corrections
by expanding $f_A^{RIA}(z,x)$ in powers of the small quantity $\Delta$.
We obtain:
\begin{equation}
f_A^{RIA}(z,x) \simeq
f_A(z) F^N_2(y) + \langle \Delta  \rangle _\perp {G}^N (y) +
{\cal O} ( \Delta^2) \,  ,
\label{expdel}
\end{equation}
where
\begin{equation}
{G}^N (y)=
\left. \frac{ \partial y'}{\partial \Delta} \right| _{\Delta=0}
\times
\left.   \frac{\partial}{\partial y'} \left[ F^N_2(y') P(y,y') \right]
\right| _{y=y'}\,  ,
\label{exd1}
\end{equation}
and
\begin{equation}
\langle \Delta \rangle _\perp =  2 \pi M_N z
\int_{k_{min}(z,\langle E \rangle_A)}^\infty
dk \, k \, n_A(k) \, \Delta
\label{deltaz}
\end{equation}
is the average value of $\Delta$
over the nucleon transverse momentum $k$.  Note that the first term
in Eq.~(\ref{expdel}) (the one proportional to
$f_A$) is identical to the nonrelativistic IA result
given in Eq.(\ref{IA}).

Expanding the first term in Eq.~(\ref{expdel}) around $z=1$, as we did
before, gives
\begin{equation}
R_A(x)-1 \simeq \alpha_1(x) \beta_1(A) + \alpha_2(x) \beta_2(A)\, ,
\label{3par}
\end{equation}
where $\alpha_1(x)$ and $\beta_1(A)$ are the $\alpha(x)$ and $\beta(A)$
given in Eq.~(\ref{ab}), and
\begin{eqnarray}
\alpha_2(x) &&=\frac{{G}^N(x)}{F_2^N(x)} \nonumber\\
\beta_2(A)&&= \langle \Delta (z) \rangle =  2 \pi M_N \int_0^\infty dz\, z
\int_{k_{min}(z,\langle E \rangle_A)}^\infty
dk \, k \, n_A(k)\, \Delta \,  .
\label{deltaA}
\end{eqnarray}
The new, relativistic correction term $\beta_2(A)$ is the value of
$\Delta$ averaged over both longitudinal and transverse momentum variables.

To summarize: the RIA still gives a result in which the $x$ dependence
of the nuclear structure function is rescaled by the motion of the nucleons,
and the $A$-dependence is still
governed by average properties of nucleon dynamics which can be readily
calculated with high accuracy by present day nuclear models.
The important difference in the region of large $x$ is that the
$A$ and $x$ dependencies of the EMC effect
cannot be written as a product of a {\it single\/} factor of a function of $x$
times a function of $A$, but require the sum of two such products.
This is a consequence of relativistic effects, which give the second term
in Eq.~(\ref{3par}).

We now turn to a discussion of the three points mentioned at the
beginning of
this letter. The shaded area in Fig.~1 are the predictions of the RIA,
including {\it an estimate of the theoretical error\/}. This error includes
uncertainties due to different parametrizations of the free nuclear
structure
functions (shown by the three dotted lines in the figure) and
variations in the nuclear parameters of our theory, namely the average
removal energy $\langle E \rangle $, and the value of the mass parameter
$m_X$ for the spectator quarks.  Of these, the only significant error
comes from the dependence of the theory on the (unknown) mass of the
spectator quarks (which is expected to be a mass close to, but larger than,
$M_N$).  The two
solid lines shown in Fig.~1 correspond to $m_X=940$ and $1800$ MeV, showing
that the predictions are insensitive to the preceise value of this
parameter.
We conclude that the agreement between theory and experiment is
no accident.

Finally, we turn to the question of the extraction of the neutron structure
function from experimental data.  The usual way to obtain the neutron
structure
function is to measure the ratio of the deuteron to proton structure
functions.
Ignoring nuclear motion effects in the deuteron, this ratio is
\begin{equation}
{F^D_2(x)\over F_{2p}(x)}-1  = {F_{2n}(x)\over F_{2p}(x)}=R(x) \, .
\end{equation}
However, if the nucleon structure functions are not modified by the nuclear
medium, and if our theory of the EMC effect is good enough, then the ratio
$R(x)$ could be, in principle, also extracted from
measurements on any other nucleus, even if the nuclear recoil effects
are large for that nucleus.  The value
of $R(x)$ obtained in this way should agree with the result obtained from the
deuteron. This idea is best tested by extracting $R(x)$ from a global fit
to all available nuclei
($A=2,4,9,12,27,40,56,108,197$) \cite{Bodeketal,Gomez} using the
theory only to specify the $A$ dependence through the functions $\beta_1(A)$
and $\beta_2(A)$.  One can then not only extract $R(x)$ from the global fit,
but also the ``coefficients'' $\alpha_1(x)$ and $\alpha_2(x)$, and hence
determine all of the $x$ dependence of the EMC effect in an independent way
which is not biased by the theory.  This idea has already been carried out in
Refs.~\cite{Bodeketal,Gomez}, where the $A$-dependence was modeled
according to an empirical formula \cite{FSPRep}.

The results of this analysis \cite{Bodeketal,Gomez}
gave rise to a puzzling situation:
while at $0.3<x<0.6$
the extracted values of $R$ were in good agreement
with the $R$ extracted from deuteron, at higher $x$ ($x>0.7$) there
appeared to be a significant discrepancy.
In particular, in the limit $x \to 1$,  the deuteron data
seems to extrapolate to the lower bound of $1/4$,
while the global fit
extrapolates to a much higher value (compatible with the $3/7$ suggested
by Farrar and Jackson \cite{Farrar}).

We have extracted the value of $R$ from a global fit to the EMC data,
using the $A$-dependence suggested by Eq.~(\ref{3par}).
The ratio is obtained by fitting the formula
\begin{eqnarray}
R^{*\, exp}_A &&=
\frac{\sigma_A^{* \, exp}(x)}{Z\,\sigma_{free}^{p \, exp}(x)} -1
= \frac{N\sigma^n_A(x) +Z\sigma^p_A(x) }{Z\, \sigma_{free}^{p \, exp}(x)}
-1 \nonumber\\ &&=
\frac{N}{Z} R(x) + \left[ 1 + \frac{N}{Z} R(x) \right]
\left[ \alpha_1(x) \beta_1(A) + \alpha_2(x) \beta_2(A) \right] \, ,
\label{eq18}
\end{eqnarray}
where the first line defines the experimental ratio $R_A^{* \, exp}$
\cite{Gomez}, and $\beta_1(A)$ and $\beta_2(A)$ are the theoretical input.
The fit determines  $R(x)$, $\alpha_1(x)$, and  $\alpha_2(x)$.
The results of the fit, shown as black squares in Fig.~2, are compared
with the theoretical results (solid curves) obtained from the
parameterizations of $F_{2p}$ and $F_{2n}$ (obtained
from deuteron data).

The open squares in Fig.~1a were obtained in Ref.~\cite{Bodeketal}
(referred to as BDR).
Note that the trend of the
BDR fit clearly suggests disagreement with the deuteron data
(the solid line)
and also suggests that $R\to 3/7$ as $x\to 1$, while our fit is full agreement
with the deuteon data, and extrapolates to a limit near $1/4$ as $x\to 1$.

It is now becoming increasingly clear that
the nature of the EMC-effect
cannot be unravelled unambigously without addressing the
problems inherent in the formalism (relativistic
covariance, meson degrees of freedom and final state interactions)
\cite{GLlong}.  Our formula includes a breakdown of factorization
due to relativistic effects.
A margin of {\em accuracy} of the theory must
be anyway specified and compared to the
magnitude of the effect.
We therefore present our results along with the theoretical
"error".  More accurate extractions could be obtained
by measuring the EMC-effect at large $x$ for a wider number of
nuclei.

It is a pleasure to thank J.Gomez and V.Gladshev for help with
and discussion of the experimental fits during the initial stages
of this work.  The support of the Department of Energy, through CEBAF, is
gratefully acknowledged.

$^*$
Permanent address: I.N.F.N., Sezione Sanita',
Viale Regina Elena 299, I-00161, Roma, Italy

\begin{figure}
\caption{The EMC data for $^{56}$Fe.  The dashed line
is the nonrelativistic calculation of CL.  The shaded area is the
relativistic calculation of GL, including an estimate
of theoretical uncertainties. }
\end{figure}

\begin{figure}
\caption{Results of a global fit to the EMC data.  The
solid squares (with error bars) are obtained from a three parameter fit
to Eq.~(\protect\ref{eq18}); the open squares (displaced slightly in
$x$ so that
they can be separated from the solid squares) are a two parameter fit with
the relativistic term $\beta_2$ omitted.  The short bars in (a) are the
limits 3/7 and 1/4.  The solid lines are discussed
in the text.}
\end{figure}

\end{document}